%% file: MAIN.tex
\documentclass[prb,twocolumn,superscriptaddress,preprintnumbers]{revtex4-2}
\usepackage[english]{babel}
\usepackage[subpreambles=true]{standalone}
\usepackage{subfiles}
\usepackage{graphicx}
\usepackage{dcolumn}
\usepackage{bm}
\usepackage{float}
\usepackage{orcidlink}
\usepackage{hyperref}
\usepackage{amsmath}
\usepackage{amssymb,latexsym,amsthm,makeidx,enumerate}
\hypersetup{colorlinks=true, linkcolor=blue, citecolor=blue, urlcolor=blue, pdftitle={Mapping of TaTe\(_{4}\) Fermi surface from SdH oscillations}, pdfauthor={Diego Silvera}}

\newcommand{\TT}{TaTe$_{4}$}

\par
\begin{document}

\title{Fermi Surface Reconstruction and Anisotropic Linear Magnetoresistance in the Charge Density Wave Topological Semimetal \TT{}}

\author{D. Silvera-Vega \orcidlink{0009-0008-5869-6545}}
\author{J. Rojas-Castillo \orcidlink{0009-0008-4051-3392}}
\affiliation{Department of Physics, {Universidad de Los Andes}, Bogotá 111711, Colombia}
\author{E. Herrera-Vasco \orcidlink{0000-0002-7319-7428}}
\affiliation{Laboratorio de Bajas Temperaturas y Altos Campos Magnéticos, Departamento de Física de la Materia Condensada, Instituto de Ciencia de Materiales Nicolás Cabrera, Condensed Matter Physics Center (IFIMAC), Facultad de Ciencias, Universidad Autónoma de Madrid, 28049 Madrid, Spain}
\author{S. Chikara \orcidlink{0000-0002-8819-4465}}
\affiliation{National High Magnetic Field Laboratory, Florida State University, Tallahassee, FL 32306, USA}
\author{E. Ramos-Rodríguez \orcidlink{0000-0002-6728-1279}}
\affiliation{Department of Physics, {Universidad de Los Andes}, Bogotá 111711, Colombia}
\affiliation{Department of Physics, {Universidad Nacional de Colombia}, Bogotá 111321, Colombia}
\author{A. F. Santander-Syro \orcidlink{0000-0003-3966-248}}
\affiliation{Université Paris-Saclay, CNRS, Institut des Sciences Moléculaires d'Orsay, 91405 Orsay, France}
\author{J. A. Galvis \orcidlink{0000-0003-2440-8273}}
\affiliation{School of Sciences and Engineering, Universidad del Rosario, Bogotá 111711, Colombia}
\author{B. Uribe \orcidlink{0000-0003-2234-4624}}
\affiliation{Departamento de Matemáticas y Estadística, Universidad del Norte, Km. 5 Vía Antigua Puerto Colombia, Barranquilla 080020, Colombia}
\author{R. González-Hernández \orcidlink{0000-0003-1761-4116}}
\affiliation{Departamento de Física y Geociencias, Universidad del Norte, Km. 5 Vía Antigua Puerto Colombia, Barranquilla 080020, Colombia}
\author{A. C. García-Castro \orcidlink{0000-0003-3379-4495}}
\affiliation{School of Physics, Universidad Industrial de Santander,
Carrera 27 Calle 09, SAN-680002, Bucaramanga, Colombia}
\author{P. Giraldo-Gallo \orcidlink{0000-0002-2482-7112}}
\affiliation{Department of Physics, {Universidad de Los Andes}, Bogotá 111711, Colombia}

\date{Received: \today, Revised: XXX, Accepted: XXX, Published: XXX}



\begin{abstract}
Understanding the interplay between topology and correlated electron states is central to the study of quantum materials. \TT{} is a quasi-one-dimensional charge density wave (CDW) compound predicted to host topological phases, which makes it a model platform to explore this interplay. Here, we combine high-field magnetotransport measurements with density functional theory calculations to provide a comprehensive mapping of the Fermi surface (FS) of TaTe$_4$ in its CDW phase. Using multiple current-field geometries, we resolve the four largest of six pockets of the FS predicted by theory and find no evidence of non-CDW bands, highlighting the full reconstruction of the FS in the bulk. We identify a previously unobserved quasi-cylindrical pocket and uncover a large size orbit consistent with magnetic breakdown between reconstructed FS sheets, from which we estimate a CDW gap of $\sim$0.29~eV. Moreover, we observe a robust linear magnetoresistance that persists across all field directions when current flows perpendicular to the 1D chains along which the CDW is formed, with a distinct high-field linear regime emerging when field is along the chains. These findings establish TaTe$_4$ as a prototypical material to study the coexistence of correlation-driven reconstruction and topological electronic states.
\end{abstract}

\maketitle

\subfile{Sections/Introduction.tex}
\subfile{Sections/Results.tex}

\subfile{Sections/Discussion.tex}

\subfile{Sections/MaterialsMethods.tex}
\bibliography{BIBLIOGRAPHY}
\subfile{Sections/Acknowledgements.tex}

\subfile{Sections/SupplementaryInfo.tex}
\end{document}

%% file: Sections/Introduction.tex
\section{\label{sec:level1} Introduction}
 
 The study of the interplay between highly correlated states, such as superconductivity or density waves and topological states, has become an increasingly active research field in the context of quantum materials due to their potential to hold novel quasi-particles \cite{RevModPhys.82.3045,RevModPhys.83.1057,Wang2016,WANG2017425,He2016,doi:10.1073/pnas.1801650115}. Topological materials are known for holding  nodal points of various types such as Weyl, Dirac and double Dirac points \cite{RevModPhys.90.015001,doi:10.1126/science.aaf5037,PhysRevLett.116.186402}. At these nodal points, new  states and phenomena could emerge when they interact with other correlated states, such as axion states or  Majorana fermions \cite{PhysRevB.87.161107,PhysRevB.83.205101,Sato_2017}. While substantial theoretical and experimental progress has been made in understanding topological phases and density waves individually,   observations of materials where topological features and electronic correlations intertwine at the level of the Fermi surface are relatively scarce.
 
Transition metal tetrachalcogenides (TMTs), i.e. TaTe\(_{4}\) and NbTe$_{4}$, are promising in this context due to their complex  band structures which include Dirac points, and a transition to a charge density wave (CDW) phase that is expected to deeply reconstruct its band structure and leads to the emergence of new high-degeneracy Dirac points  \cite{PhysRevB.108.155106,doi.org/10.1002/inf2.12156,PhysRevB.102.035125,PhysRevB.110.125151,dhva_NbTe4}. Therefore, these materials stand out as ideal systems for probing the interplay between topological features and CDW order.

For the case of \TT{} there is consensus on the fact that it is a CDW-compound at room temperature, but its Fermi surface (FS) has not been fully resolved using experimental approaches. Recent angle-resolved photoemission spectroscopy (ARPES) experiments have shown that TaTe\(_{4}\) has a dominant feature consistent with the periodicity of its non-CDW Brillouin zone, along with an additional weaker feature with reduced spectral weight that reflects the CDW periodicity\cite{PhysRevB.110.125151, PhysRevB.108.155121}. Other reports suggest that certain bands are more affected by the CDW transition than others, suggesting an orbital-selective CDW reconstruction in TaTe\(_{4}\) \cite{Xu2023}, which might explain the existence of pockets or bands with different periodicities. More recently, through a combined theoretical and ARPES approach \cite{PhysRevLett.133.116403}, the observed features of the band structure have been linked to a CDW-induced band folding. Despite these advances, ARPES has so far failed to reveal the complete FS of \TT{}
likely due to its surface sensitivity, limited resolution, and matrix element effects. These advances highlight the need for complementary, bulk-sensitive techniques to uncover the dominant pockets and the overall \TT{} FS geometry.

Alongside ARPES, transport measurements have been employed to explore the intricate features of the FS in TMTs \cite{10.1063/1.5005907}. The Shubnikov-de Haas (SdH) effect has been used to relate extremal orbit areas in k-space at the Fermi level under a magnetic field to oscillations in  the electrical resistance of TaTe\(_{4}\). By isolating each of the frequencies contributing to those oscillations, it is possible to obtain all  FS sections of a material, even sections with very small cross-sectional areas. However, previous SdH studies on \TT{}
have only identified limited information of the experimental FS: few pockets, most in a limited k-space region;  therefore very limited information has so far been revealed through this bulk probe \cite{PhysRevB.102.035125,2017ApPhL.110i2401L}.

In this paper, we present a detailed band structure analysis of TaTe\(_{4}\) using a combination of DFT calculations and high-field magnetoresistance measurements, providing detailed insights into the FS, not addressed in previous works. We report magnetoresistance data for TaTe\(_{4}\) under magnetic fields up to 35 T oriented and rotated along two perpendicular planes, offering a clear image of the Fermi surface of the material. We found that the bulk Fermi surface of this material fully matches the DFT predictions for the \textit{CDW-phase} with four of six predicted bands identified. No evidence of non-CDW FS was found. In addition, we show that a linear magnetoresistance appears at all angles at low magnetic fields when imposing a current perpendicular to the 1D chains (this is, along the $a$-axis). Furthermore, we report the presence of a high Shubnikov-de Haas (SdH) frequency in the magnetoresistance data at certain angles in this configuration, which is consistent with magnetic breakdown effect,and that a secondary linear magnetoresistance regime appears at high magnetic fields at angles for which magnetic breakdown occurs. Our work provides complete and detailed information of the band structure of this material at the Fermi level, which is paramount in the understanding of the possible connection between topological and correlated phases of matter.

\begin{figure*}[!t]
        \centering
         \includegraphics[width=\textwidth]{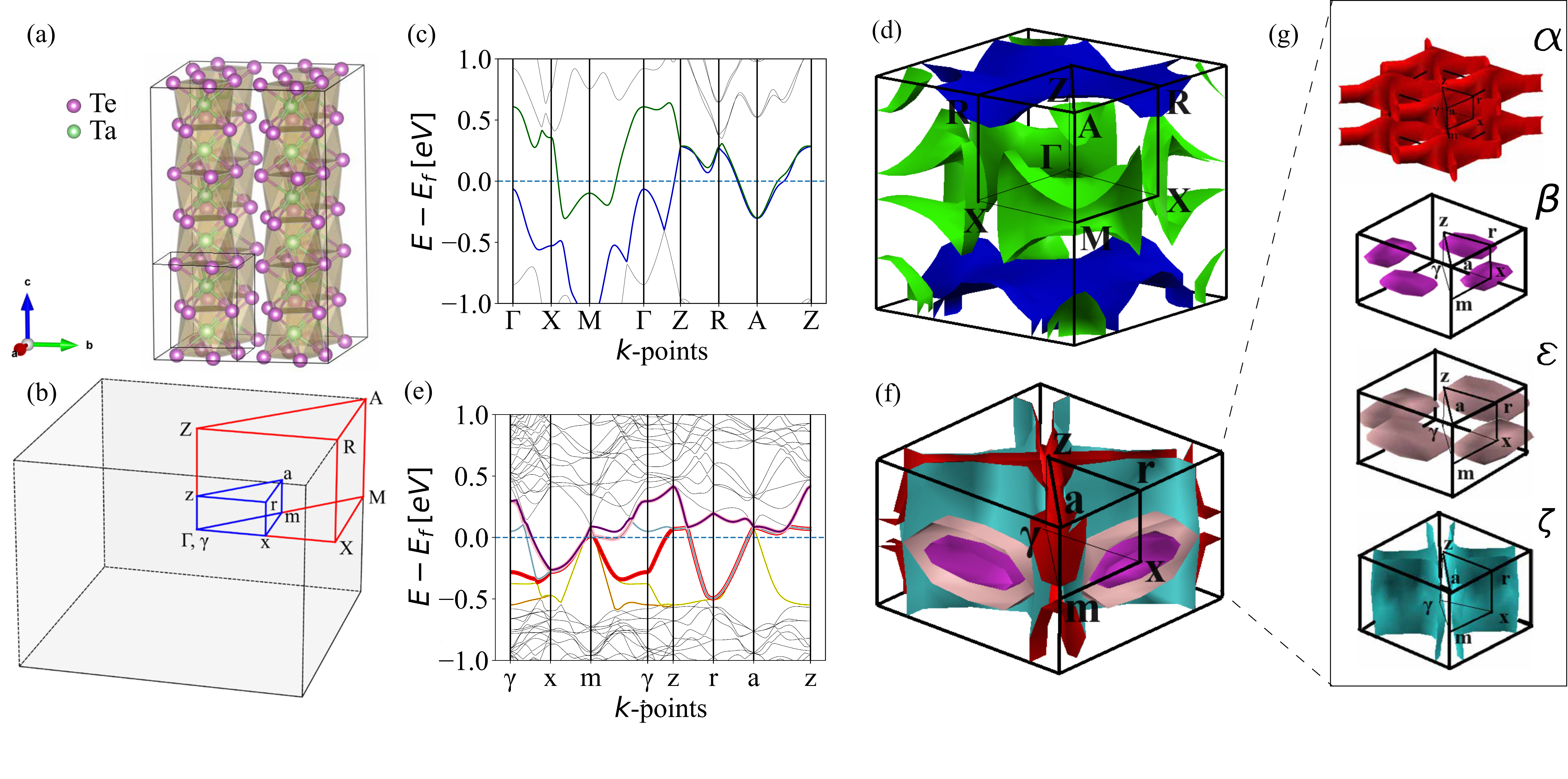}
        \caption{(Color online) (a) Unit cell of TaTe\(_{4}\) in both the CDW (large rectangular prism) and non-CDW (small rectangular prism) phases. (b) High symmetry points of the first brillouin zone of TaTe\(_{4}\), in both its high symmetry non-CDW phase (corners of red polygon with capital letters) and low symmetry CDW phase (corners of red polygon with lowercase letters). (c-d) and (e-f) are the band structure and Fermi surface of \TT{} obtained through DFT calculations for the non-CDW phase and the CDW phase, respectively. In (c,e) bands crossing the Fermi level are highlighted in colors other than black. (g) Four of the CDW FS pockets shown individually and in the corresponding band color of (e) to better illustrate the possible closed orbits for each. The other two FS pockets coming from bands in yellow and orange in (e), and centered around $a$ and $m$ points, are too small to be visualized.}
        \label{fig:intro}
    \end{figure*}

%% file: Sections/Results.tex
\section{Results and discussion}
\subsection{Band structure calculations}

TaTe$_4$ crystallizes at high temperature in a $P4/mcc$ tetragonal unit cell, and is a well-known quasi-1D material. It is composed of chains of Ta atoms in the center of two Te-square antiprisms, extending along the crystallographic \textit{c}-axis. These chains are coupled to others through weak van der Waals forces, which explains the quasi-1D character of this material (Fig. \ref{fig:intro}(a)). This phase corresponds to the non-CDW phase, which unit cell is depicted in the small rectangle of Fig. \ref{fig:intro}(a). Its corresponding first Brillouin zone high symmetry points are shown in the corners of the red polyhedron of Fig. \ref{fig:intro}(b), and labeled in capital letters.  
In this phase the electronic structure reveals a weak-metallic behavior with clear topological features highlighted by the Dirac nodal points located  between the $\Gamma$ -- Z path, as in Fig. \ref{fig:intro}(c). Thus, the high-symmetry phase falls into the enforced semimetal with Fermi degeneracy (ESFD) categorization \cite{doi:10.1126/science.abg9094}. The non-CDW Fermi surface exhibits several closed surfaces centered on the M and A points and quasi-flat bands parallel to the Z plane, as shown in Fig. \ref{fig:intro}(d).

Once the temperature is lowered below about 400 K, the $P4/mcc$ to the $P4/ncc$ displacive phase transition takes place and the CDW phase is obtained \cite{PhysRevB.107.045120}, with a new unit cell $2a \times2a\times3c$ (large rectangular prism in Fig. \ref{fig:intro}(a)) due to a A-B-A stacking between adjacent chains. Its new Brillouin zone high symmetry points are shown in the corners of the blue polyhedron of Fig. \ref{fig:intro}(b), and labeled in lowercase letters. With this, the band structure is drastically modified, opening gaps at the points where the non-CDW Dirac points appeared and leading to the emergence of new high-degeneracy points, including an eightfold-degenerate crossing at point $a$ which has been indexed as a Double Dirac point in previous works \cite{PhysRevB.102.035125}. Six bands cross the fermi level (Fig. \ref{fig:intro}(e)), leading to the FS portrayed at Fig. \ref{fig:intro}(f). It consists of one hole-like band labeled as $\alpha$ in Fig. \ref{fig:intro}(g), which extends along the $z$-$r$-$a$ plane of the CDW Brilloiun zone and contains a combination of closed and open orbits; two electron-like semi-ellipsoidal pockets centered at the x-point, labeled as $\beta$ and $\epsilon$; and one electron-like cylindrical pocket dubbed as $\zeta$, with its main axis centered at $x$ and extending along the $x$-$r$ direction. Two additional bands cross the Fermi level leading to very small hole orbits centered at $m$ and $a$ high symmetry points. However, due to their small size and the limited $k$-space grid used for the interpolation, it is difficult to visualize them. Interestingly, these pockets are also not observed in our magnetoresistance measurements, suggesting that their small size is indeed real.

\subsection{Magnetoresistance measurements}

\subsubsection{Measurement configurations}

\begin{figure*}[!t]
\includegraphics[width=\textwidth]{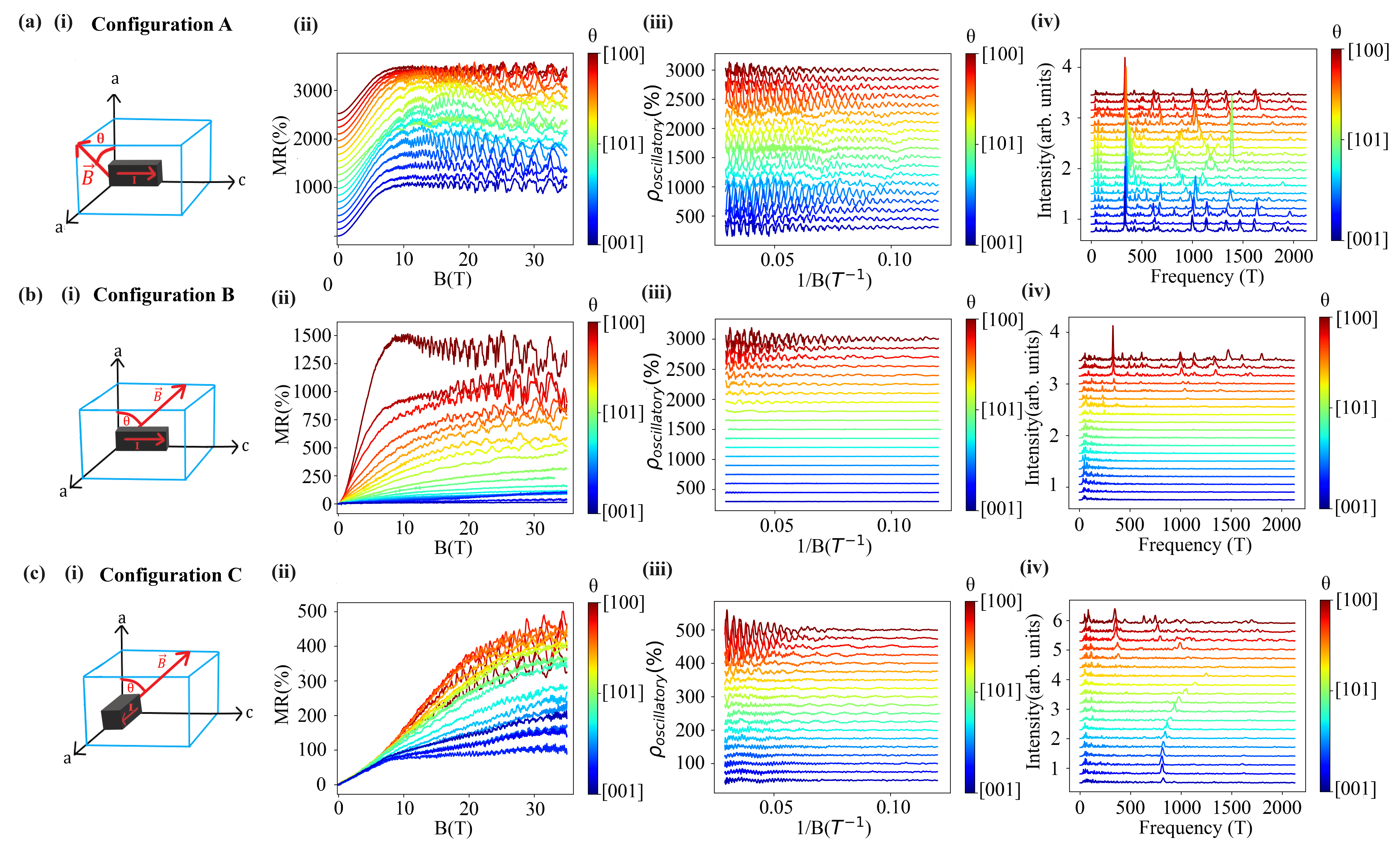}
\caption{(Color online) Magnetoresistance curves for magnetic field at different rotation planes and current directions. From top to bottom: Row (a) corresponds to rotation of magnetic field within the \textit{a}-\textit{a} plane, with current along \textit{c} (configuration A); row (b)  corresponds to rotation of field within the \textit{a}-\textit{c} plane, with current along \textit{c} (configuration B); and row (c)  corresponds to the magnetic field rotation  within the \textit{a}-\textit{c} plane, but with current along \textit{a} (configuration C). From left to right: Column (i) for each row depicts the magnetic field and current setup for each case; column (ii) shows the symmetrized magnetoresistance data for all angles for each configuration as a function of magnetic field. The color scale represents the $\theta$ angle of orientation of the magnetic field with respect to the crystalline axis. For config. A an offset for each curve was added for better visualization of the plots; Column (iii) shows the oscillatory component of magnetoresistance curves as a function of the inverse of magnetic field; and column (iv) presents the fast Fourier transform of their corresponding curves in (iii), highlighting their dominant frequency composition and their evolution with field orientation.}
\label{fig:MR3configurations}
\end{figure*}

Due to the quasi-1D character  of the crystalline structure of \TT{}, crystal growth usually results in very thin rectangular parallelepiped crystals, with an elongated \textit{c}-axis. Because of this morphology, four-probe electric contacts are therefore usually placed such that current runs along this crystallographic orientation. In our case, flux-method grown crystals exhibited this morphology, while crystals synthesized through the CVT method appeared to have a more isotropic aspect-ratio, as mentioned in the previous section. This was advantageous for the purposes of this work, as it allowed us to take four-probe transport measurements with current along the \textit{a}-axis, in addition to the typical $c$-axis. Transport measurements were performed in the presence of an applied external magnetic field  along different orientations within two high- symmetry crystallographic planes. By combining different current directions and magnetic field rotation planes, we implemented three distinct configurations, labeled A through C, each providing complementary information about the electronic structure of \TT{}. 
In configuration A, current flows along the \textit{c}-axis while the magnetic field rotates in the \textit{a}-\textit{a} plane, ensuring the field remains perpendicular to the current (Fig. \ref{fig:MR3configurations}a(i)). In contrast, in configuration B current also flows along \textit{c}, but the field rotates in the \textit{a}-\textit{c} plane, ending aligned parallel to the current (Fig. \ref{fig:MR3configurations}b(i)). 
While both configurations had been previously used in magnetotransport studies on \TT{} \cite{10.1063/1.5005907,PhysRevB.102.035125}, they were limited to lower magnetic field magnitudes. The strong suppression of magnetoresistance when current and field were aligned along the \textit{c}-axis (Fig. \ref{fig:MR3configurations}b(ii)) motivated the introduction of configuration C. In this configuration, CVT-grown crystals were used to enable current flow along the \textit{a}-axis, while the magnetic field was rotated within the \textit{a}-\textit{c} plane (Fig. \ref{fig:MR3configurations}c(i)). As in configuration A, the current and magnetic field remained perpendicular throughout the rotation.

\begin{figure*}[!t]
        \centering
        \includegraphics[width=\textwidth]{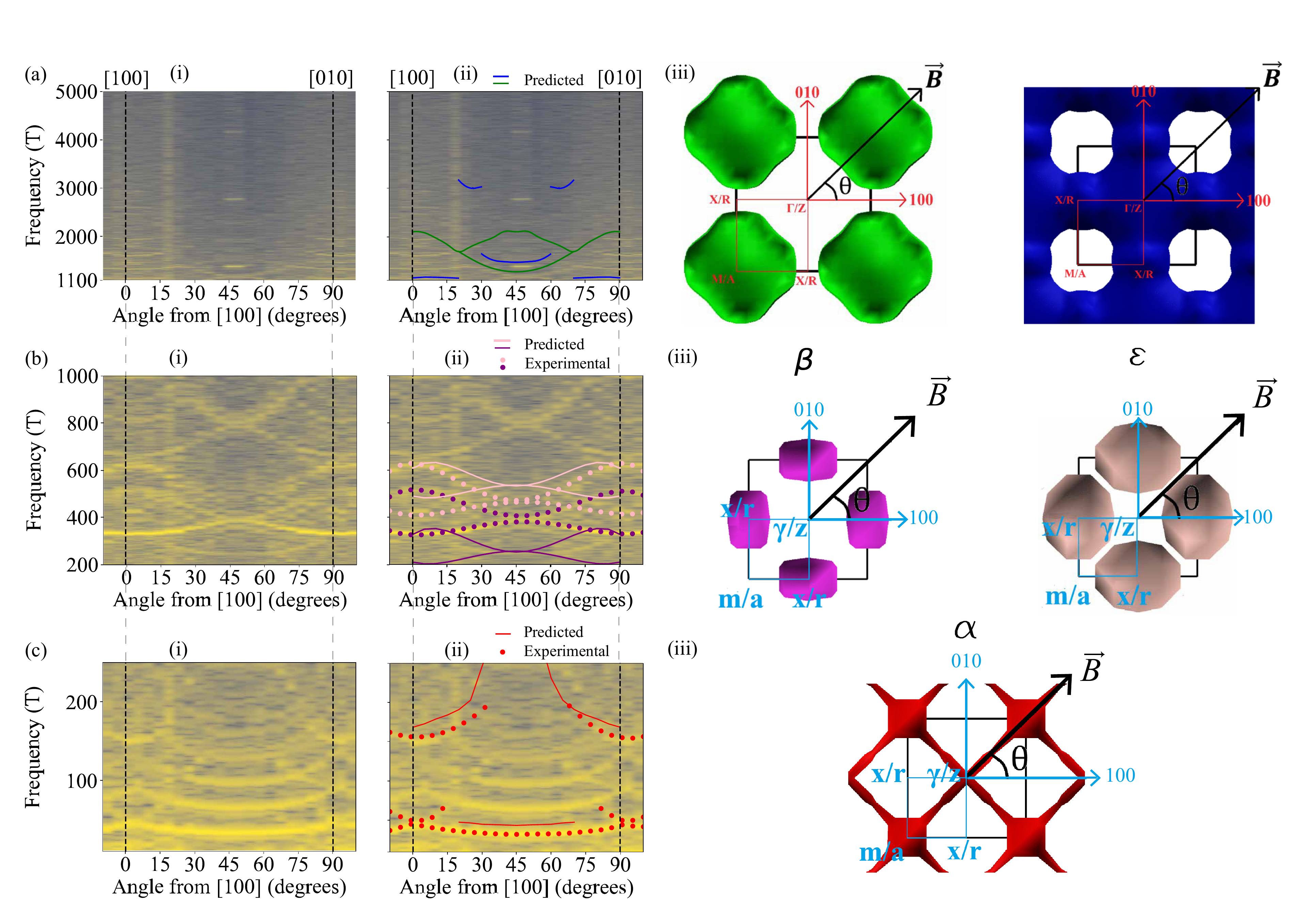}
        \caption{(Color online) Evolution of  extremal areas ($i.e.$ frequencies) of pockets for different orientations of the magnetic field in configuration A. Rows (a), (b) and (c) correspond to high, mid, and low frequency regimes, respectively. Column (i) shows contour plots of the FFT in Fig. 2(a.iv) to highlight the evolution of frequencies as a function of the magnetic field orientation. Yellow color represents a high intensity, and dark blue represent zero intensity. Column (ii) shows the same contour plots now identifying the dominant frequency branches (solid dots), and comparing them with the DFT-predicted evolution of extremal areas of pockets (solid lines). Column (iii) shows the DFT pockets that explain the evolution of the observed frequencies. The color of each pocket is chosen to match the color of the dotted and solid lines in column (ii).}
        \label{fig:MR2}
    \end{figure*}

\begin{figure*}[!t]
\includegraphics[width=\textwidth]{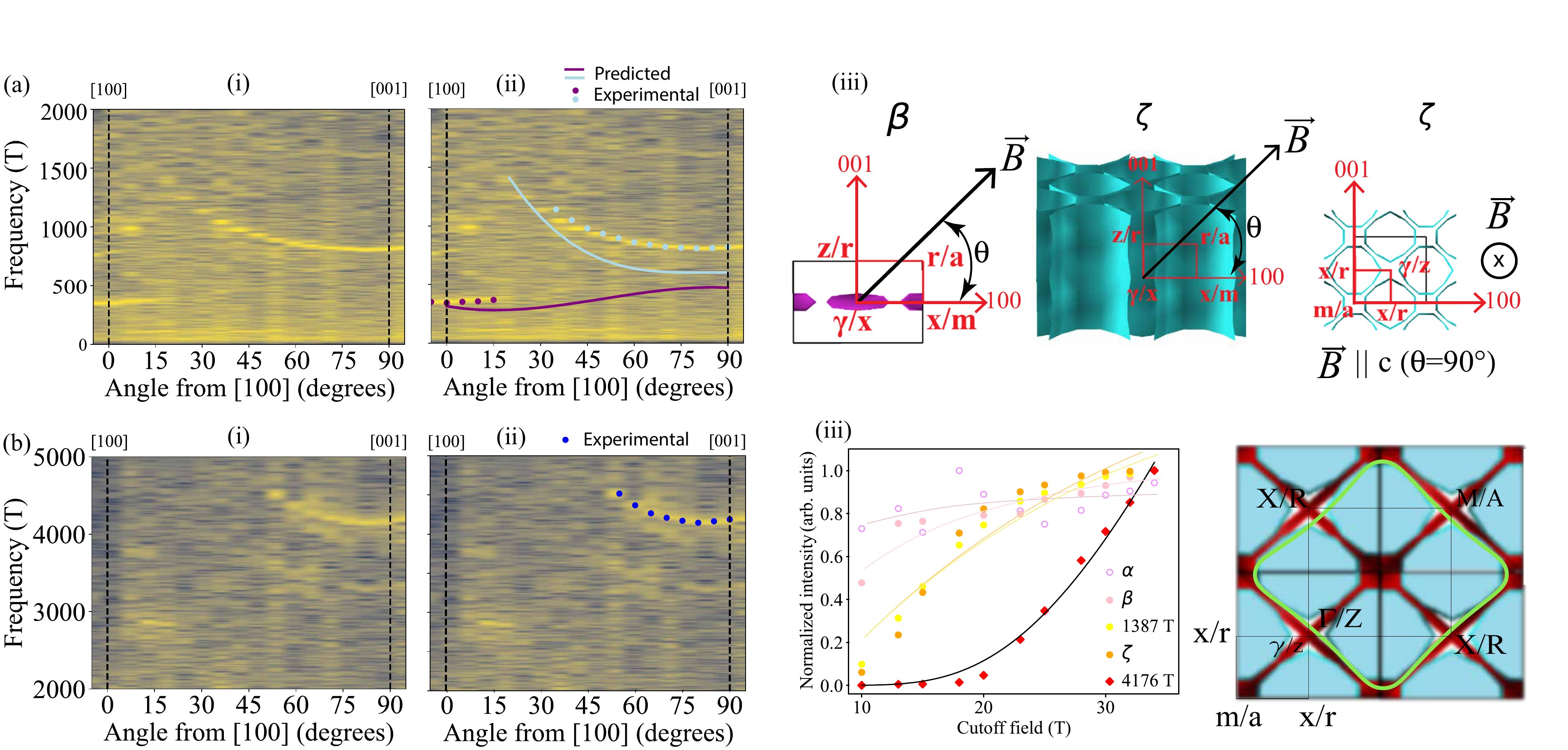}
\caption{\label{fig:FFT2}(Color online) Evolution of extremal areas ($i.e.$ frequencies) of pockets for different orientations of the magnetic field in configuration C. Rows (a) and (b) correspond low/mid and high frequency regimes, respectively. Column (i) shows contour plots of the FFT in Fig. 2(c.iv) to highlight the evolution of frequencies as a function of the magnetic field orientation; Column (ii) shows the same contour plots now identifying the dominant frequency branches (solid dots), and comparing them with the DFT-predicted evolution of extremal areas of pockets (solid lines); Figures in (a)(iii) show the DFT pockets that explain the evolution of the observed frequencies, and the color of each pocket is chosen to match the color of the dotted and solid lines in the previous column. Figures in (b)(iii) show, in the left, the evolution of the amplitude of the FFT peaks associated to different frequencies (filled dots) as a function of the lower-cutoff field to perform the FFT. Solid lines represent fits to a $e^{-B_0/B}$ form. In the right, an schematic unfolding of the CDW bands into the non-CDW Brilloin zone is made, showing in green a putative magnetic breakdown orbit that could originate the $\sim$ 4 kT (at 90°) frequency seen in the data.}
\end{figure*}

Resistance measurements were performed for each of the three configurations under magnetic fields between -35 T and + 35 T, at 1.3 K.  To isolate the longitudinal magnetoresistance and to eliminate possible contributions from the Hall component, the data was symmetrized with respect to the magnetic field (Figs. \ref{fig:MR3configurations}(a,b,c)(ii)). In every configuration, magnetoresistance exhibited both a non-oscillatory and an oscillatory component, the latter as a consequence of the SdH effect. Nevertheless, all three configurations displayed different features, revealing the anisotropic nature of the FS of this material. 

In configuration A, a large magnetoresistance is measured, which saturates to about 1400 $\%$ for certain magnetic field orientations as shown in Fig. \ref{fig:MR3configurations}(a)(ii). The SdH oscillations are appreciable from magnetic fields as low as 5 T. On the other hand, configuration B shows that both non-oscillatory and oscillatory components are suppressed as the magnetic field approaches the \textit{c}-axis, i.e., the current direction, similar to the results in \cite{10.1063/1.5005907}. In contrast, in configuration C both the oscillatory and non-oscillatory components of magnetoresistance could be observed at all angles, even when the magnetic field was aligned with the \text{c}-axis. This configuration thus provides access to additional features of the Fermi surface.  Notably, we observe that there is a strong linear contribution to the non-oscillatory magnetoresistance across all field orientations in this case. Moreover, when the magnetic field is close to the \textit{c}-axis, the non-oscillatory component of the magnetoresistance displayed a secondary linear regime, with a reduced slope. This linear magnetoresistance is accompanied by an oscillatory component with a very large frequency that is absent at other magnetic field orientations. In a subsequent section we will discuss the possible causes of both the linear magnetoresistance in \TT{} and this high frequency component.

\subsubsection{Oscillatory component of the magnetoresistance}

In order to retrieve information from the FS of \TT{}, magnetoresistance measurements were processed to isolate their oscillatory component, which arises from the SdH effect. This was achieved by substracting a smooth background through a spline fit, yielding the oscillatory traces in the plots shown in Figs. \ref{fig:MR3configurations}(a,b,c)(iii). After that, a Fast Fourier transform (FFT) was applied to the oscillatory signal vs inverse field data, and the dominant frequencies were retrieved for all magnetic field orientations in each configuration, as presented in Figs. \ref{fig:MR3configurations}(a,b,c)(iv)). From these frequencies the extremal cross-sectional areas of the Fermi surface ($A_{FS}$) of \TT{} were determined following the Onsager relation $F_{1/B}=(\phi_0/2\pi^2)A_{FS}$, where $\phi_0=$2.07$\times$10$^{-15}$ T m$^2$ is the magnetic flux quantum. 
Due to the presence of multiple oscillation frequencies, each corresponding to different sections or \textit{pockets} of the FS of the material, the analysis of the evolution of such frequencies with the magnetic field orientation was conducted using the color map plots of Figs. \ref{fig:MR2}(a,b,c)(i) and (ii). In those plots, the colors represent how dominant each frequency is. Intense yellow indicates dominant frequencies, while dark blue represent frequencies that do not contribute to the oscillatory signal. In order to facilitate the analysis, the frequency space was divided in three different regimes: a high frequency regime, from 1000 T to 5000 T, a mid-frequency regime, from 300 T to 1000 T, and a low frequency regime, below 300 T. 

The analysis for different frequency regimes was motivated by recent ARPES measurements \cite{PhysRevLett.133.116403, PhysRevB.110.125151,PhysRevB.108.155121}. These studies revealed that the most intense spectral weight at the Fermi level are ellipsoids centered at the X/M-point in reciprocal space which follow the periodicity of the high-symmetry non-CDW phase, with a cross  sectional area about ~0.4 $\AA^{-2}$.  According to the Onsager relation, such pocket would correspond to a frequency close to 4000 T when field is perpendicular to the \textit{a}-\textit{c} plane, which corresponds to  0° and 90° in configuration A (Figs. \ref{fig:MR2}(a)(i,ii)) and 0° for configurations B and C (Figs. \ref{fig:FFT2}(a)(i,ii)). However, no frequency component is observed around the \textit{a}-direction in the high frequency regime in any configuration. Therefore, there is no evidence of such pocket from the SdH measurements. In configuration A and for this high frequency range, the only contribution occurs in a limited range of field orientations close to 45°, where frequencies of 1387 T $\pm$ 8 T, 2774 T $\pm$ 11 T and 4163 T $\pm$ 15 T  are detected (see Fig. \ref{fig:MR2} (a)(i)). These contributions correspond to the fundamental, first, and second harmonics of the 1387 T signal. Although this frequency value lies in the range of closed orbits that could originate in the FS of the non-CDW phase, the angle dependence of this signal does not agree with that of the predicted non-CDW pockets, which are displayed in Fig. \ref{fig:MR2}(a)(iii).

Figures \ref{fig:MR2}(b)(i,ii) show the angular evolution of the oscillation frequencies in configuration A within the mid-frequency regime (200 T to 1000 T). This range contains the most prominent contributions to the oscillatory magnetoresistance in this configuration. The purple and light-pink dots in Fig. \ref{fig:MR2}(b)(ii) trace the angle dependence of the fundamental frequencies of the four most dominant branches observed in the data. Their angular evolution closely resembles the expected behavior of the calculated $\beta$ and $\epsilon$ electron pockets of the CDW-phase Fermi surface, indicated by the solid purple and light-pink lines in the same figure. While the predicted pocket sizes differ slightly—an expected deviation in DFT calculations—the qualitative agreement supports this assignment. These four dominant branches therefore originate from two sets of nearly ellipsoidal FS pockets, centered at the same symmetry point (x) in reciprocal space and with similar sizes, as illustrated in Fig.~\ref{fig:MR2}(b)(iii). All other observed frequency branches in this regime can be indexed as higher harmonics of the $\beta$ and $\epsilon$ pockets. In fact, oscillations up to the fourth harmonic of the $\beta$ pocket can be resolved in the data, highlighting both the high quality of the crystals and the high resolution of the measurements.

Since $\beta$ and $\epsilon$ frequencies are the most dominant in the FFT spectrum, it was possible to track their evolution with temperature and calculate their effective masses. To estimate the contribution of
charge carriers to the electronic properties of TaTe\(_{4}\), the Liftshitz-Kosevich (LK) formula for the oscillatory magnetoresistance $\tilde\rho(B)$, as presented in equation 1 of SI, is used \cite{shoenberg_1984}. By fitting the evolution of the oscillatory magnetoresistance with temperature as predicted by this equation, the effective cyclotron masses of each pocket for a fixed angle, \(m^{cyc}_{\theta}\), can be obtained. In our case, the effective masses were estimated by filtering the oscillatory signal of each pocket with a band-pass filter to select a single frequency, resulting in a better fit to our data than by performing the fit on the sum of all frequencies with the whole LK formula. 

Using this approach and following the magnetic field orientations defined for configuration A in previous sections, we obtained \(m^{cyc}_{\beta-0^{\circ}}=(0.30\pm 0.03)m_{e}\) for the lower frequency branch of pocket \(\beta\) at 0\(^{\circ}\)  and \(m^{cyc}_{\beta-45^{\circ}}=(0.32\pm 0.03) m_{e}\) at 45\(^{\circ}\) in configuration A (see SI and Fig. S1 for more details), which suggests that the mass of charge carriers in these pockets is highly isotropic along this rotation plane. 

The angular evolution of low-frequency oscillations in configuration A is shown in Figs. \ref{fig:MR2}(c)(i,ii). A clear dominant frequency branch is observed, evolving from 47 T~$\pm$2 T at 0° and 90° to 30 T$\pm$3 T around 45°, as highlighted by the red dots in Fig.\ref{fig:MR2}(c)(ii). This branch remains visible up to the fourth harmonic. Its angular dependence and magnitude are consistent with the cylindrical sections of the $\alpha$ pocket predicted for the CDW-phase Fermi surface (see Figs. \ref{fig:intro}(g) and \ref{fig:MR2}(c)(iii)), as emphasized by its comparison to the predicted evolution of this orbit (lower red solid line in Fig.~\ref{fig:MR2}(c)(ii)). The minimum frequency occurring around 45° supports the prediction that these sections are elongated cylinders aligned along the \textit{z}-\textit{a} direction. A second contribution, arising from the wider sections of the same FS pocket centered at the \textit{a}-point of the CDW Brillouin zone appears near 150 T at 0° and increases rapidly as the magnetic field is rotated toward 45°. This behavior suggests that both the elongated and wider sections are connected, forming a single, merged FS pocket, as expected from our DFT calculations.

The previous identification has implications for the high-frequency signal observed at 1387 T in the high-field regime of configuration A. The rapid increase in extremal area near 45° seen in the low-frequency regime suggests that this high-frequency component could originate from closed orbits within the  FS $\alpha$ pocket, rather than from magnetic breakdown. Although magnetic breakdown had  been previously proposed to explain this feature \cite{PhysRevB.102.035125}, the fact that the 1387 T oscillation is already visible at fields as low as 8~T challenges that interpretation. Given that the probability of magnetic breakdown dependens with magnetic field as $e^{-B_0/B}$, where $B_0$ is the breakdown field, a large intensity for this contribution is unlikely at such fields.

The \(\alpha\) pocket is anisotropic in terms of effective mass and shape. Its charge carriers effective mass smoothly changes with angle, with \(m^{cyc}_{\alpha-0^{\circ}}=(0.21 \pm 0.03)m_{e}\) at 0\(^{\circ}\) and \(m^{cycl}_{\alpha-45^{\circ}}=(0.10 \pm 0.02) m_{e}\) at 45\(^{\circ}\) in configuration A. Further details of the effective mass analysis can be found in SI.

Rotation of the magnetic field within a secondary, high-symmetry plane, the \textit{a}-\textit{c} plane, was performed in configurations B and C. As previously mentioned, both non-oscillatory and oscillatory components of the magnetoresistance are strongly suppressed in configuration B for magnetic field directions close to the \textit{c}-axis, for which current and magnetic field are parallel. This outcome is expected: since the Lorentz force vanishes when current and field are aligned, the non-oscillatory part of the magnetoresistance is suppressed. Furthermore, given that the oscillatory component of magnetoresistance ($\tilde{\rho}$) is a fraction of the non-oscillatory component ($\rho_0$), as predicted by Liftshitz-Kosevich theory ($(\tilde{\rho}-\rho_0)/\rho_0\propto\cos(2\pi (F_{1/B}/B)-\phi$), the oscillatory component will also be suppressed when current and magnetic field are parallel. 

In contrast to configuration B, configuration C, for which the magnetic field is always perpendicular to the current, exhibited a clear evolution of different dominant oscillation frequencies at all angles in the $a-c$ plane, as shown in Figs. \ref{fig:FFT2}(a)(i) and (b)(i). In this case, we defined two different frequency regimes for the analysis: a low/mid frequency regime below 2000 T and a high frequency regime, from 2000 T to 5000 T. As in configuration A, the low/mid-frequency regimes dominated the oscillations in configuration C (Fig. \ref{fig:FFT2}(a)(i)). The evolution of the $\beta$ pocket, already clearly identified in configuration A, could be tracked at angles near the \textit{a}-axis; however, its frequency gradually increases and eventually disappears as the magnetic field orientation approaches the \textit{c}-axis — an effect also observed in configuration B. In contrast, as the $\beta$ pocket  oscillation faded, another frequency of oscillation appeared in the range of  1040 T $\pm$ 8 T at 35° and 829 T $\pm$ 7 T at 90°. This frequency follows a clear cylinder-like angular dependence following $1/\cos(\pi/4-\theta)$, in agreement with the DFT predicted $\zeta$ pocket (Figs. \ref{fig:FFT2}(a)(ii,iii)). This feature had not been reported in previous magnetotransport studies, likely due to the experimental challenges of imposing current along the \textit{a}-axis in \TT{} derived from the quasi-1D morphology of the crystals, as discussed earlier.

\subsubsection{High frequency regime and magnetic breakdown effects}

In the high-frequency regime of configuration C, we were able to track a prominent frequency evolving from 4176 T ± 20 T at 90° to 4520 T ± 13 T near 50°, following an angular dependence close to \(1/\cos(\pi/4 - \theta)\), consistent with a quasi-cylindrical Fermi surface (Fig. \ref{fig:FFT2}(a)(ii)). This frequency branch fades as the magnetic field orientation approaches the \textit{a}-axis. Due to its magnitude, this frequency cannot be assigned to any of the predicted CDW FS pockets. Therefore, it is worth considering whether this signal arises from cylindrical sections within the non CDW Brillouin zone. However, while the non-CDW phase does feature a quasi-2D pocket with circular cross sections perpendicular to the \textit{c}-axis, these would yield frequencies around 2000 T, significantly lower than the observed value, and expected to be observed in a more limited angle range.

Another possibility is that this signal represents a harmonic of a known pocket. Given its similar angular dependence, it is natural to speculate on a connection to the \(\zeta\) pocket. However, this is not the case : the observed 4176 T frequency is not an  multiple integer of the \(\zeta\) frequency (829 T), and its amplitude is larger than the one of the fundamental $\zeta$ pocket. Thus, this large, unindexed frequency remains inconsistent with known fundamental or harmonic oscillations from the CDW phase pockets.

Since this large frequency does not arise from neither the non-CDW phase nor a harmonic frequency of a CDW pocket, magnetic breakdown emerges as a feasible origin. This phenomenon occurs when charge carriers tunnel across small energy gaps between adjacent FS pockets, usually close to Bragg planes in reciprocal space, under the influence of  a magnetic field. In fact, the condition for easy breakdown reads as $\hbar\omega_c\geq \frac{E_{gap}^{2}}{E_F}$, where $E_{F}$ is the Fermi energy, $E_{gap}$ is a gap energy, and $\omega_c=\frac{eB}{m}$ is the cyclotron frequency.  The gap represents an energy difference between two FS sections involved in breakdown, which could be related to a reciprocal space distance $\Delta k$ through the dispersion relation as $\Delta k\sim \frac{E_{gap}}{|\text{d}\epsilon/\text{d} k|}$. The previous condition indicates that the higher the magnetic field, the larger the probability of breakdown.  Due to the fact that the breakdown probability evolves exponentially with the field, as $e^{-\frac{B_0}{B}}$, with $B_0=\frac{mE^2_{gap}}{\hbar e E_{F}}$, testing whether the intensity of the FFT evolves following this trend might help in elucidating the origin of magnetic breakdown of this frequency. However, the amplitude of the SdH oscillations also depends on an $e^{-\frac{B_{0}}{B}}$ term, as given by the Lifshitz-Kosevich formula. Within this framework, the amplitude of oscillations is modulated by the Dingle damping term that follows the previous exponential form, with $B_{0}=(14.7 \text{ T/K})(m_{cycl}/m_e)T_D$, where $T_D$ is the Dingle temperature and $m_{cycl}$ is the cyclotron mass of the pocket. This exponential dependence is inherent to all quantum oscillations and provides an alternative explanation for the rapid increase of amplitude of oscillations with field. To distinguish between these two possibilities, we compare in Fig. \ref{fig:FFT2}(a)(iii) the field dependence of the magnetic breakdown candidate frequency with that of other lower-frequency oscillations corresponding to well-identified FS pockets. While the intensity of all frequencies  displays an exponential increase with field, consistent with $e^{-B_{0}/B}$, the intensities of the CDW-indexed pockets tend to saturate at high fields, implying their effective $B_{0}$ values are much smaller in comparison. If for this large unindexed frequency the intensity increase was due solely to Dingle damping, it would imply a significantly larger Dingle temperature (i.e. lower effective scattering rate) than for all the CDW pocket, which seems highly unlikely.

The intensity of this large unindexed frequency is plotted against cutoff field and fitted to the exponential form in Fig. \ref{fig:FFT2}(a)(iii), obtaining  $B_0=106.85$ T. Furthermore, from the condition for easy breakdown, and setting $E_{F}=6.66$ eV, which is the value of the Fermi energy from our DFT calculations,  the energy gap  $E_{gap}$ can be estimated. Interestingly, the energy gap obtained is $E_{gap}$=0.29 eV, remarkably close to the energy below $E_F$ at which ARPES measurements show a significant depletion in spectral weight, and assigned to the CDW gap. These observations support the interpretation that the observed high-frequency oscillation arises from magnetic breakdown occurring at Bragg planes where a CDW-induced gap opens.

Further insight can be obtained by conceptually extending the CDW Brillouin zone to the periodicity of the high-symmetry non-CDW phase, as schematically shown in the right-hand side of Fig. \ref{fig:FFT2}(a)(iii). In this representation, the $\alpha$ and $\zeta$ pockets (red and light-blue sections, respectively) lie in close proximity in momentum space, which could enable tunneling of charge carriers between them. In particular, a magnetic breakdown orbit that connects the cylindrical section of the $\zeta$ pocket with the tubular portions of the $\alpha$ pocket could give rise to a large, composite, cylinder-like orbit—consistent in both shape and extremal area with the observed 4176 T frequency.

The results discussed in the previous two subsections have provided evidence of four out of the six predicted pockets for the CDW-phase FS, with no evidence of contributions coming from bands remanent from the non-CDW high symmetry phase, in contrast to results by recent ARPES works \cite{PhysRevB.108.155121,PhysRevB.110.125151,Xu2023}.
Therefore we can conclude that the bulk transport band structure of this material is dominated by the CDW-reconstructed bands, which closely match the size and shape of the FS predicted by DFT.

%% file: Sections/Discussion.tex

\subsubsection{Linear magnetoresistance}

\begin{figure*}[!t]
        \centering
        \includegraphics[width=\textwidth]{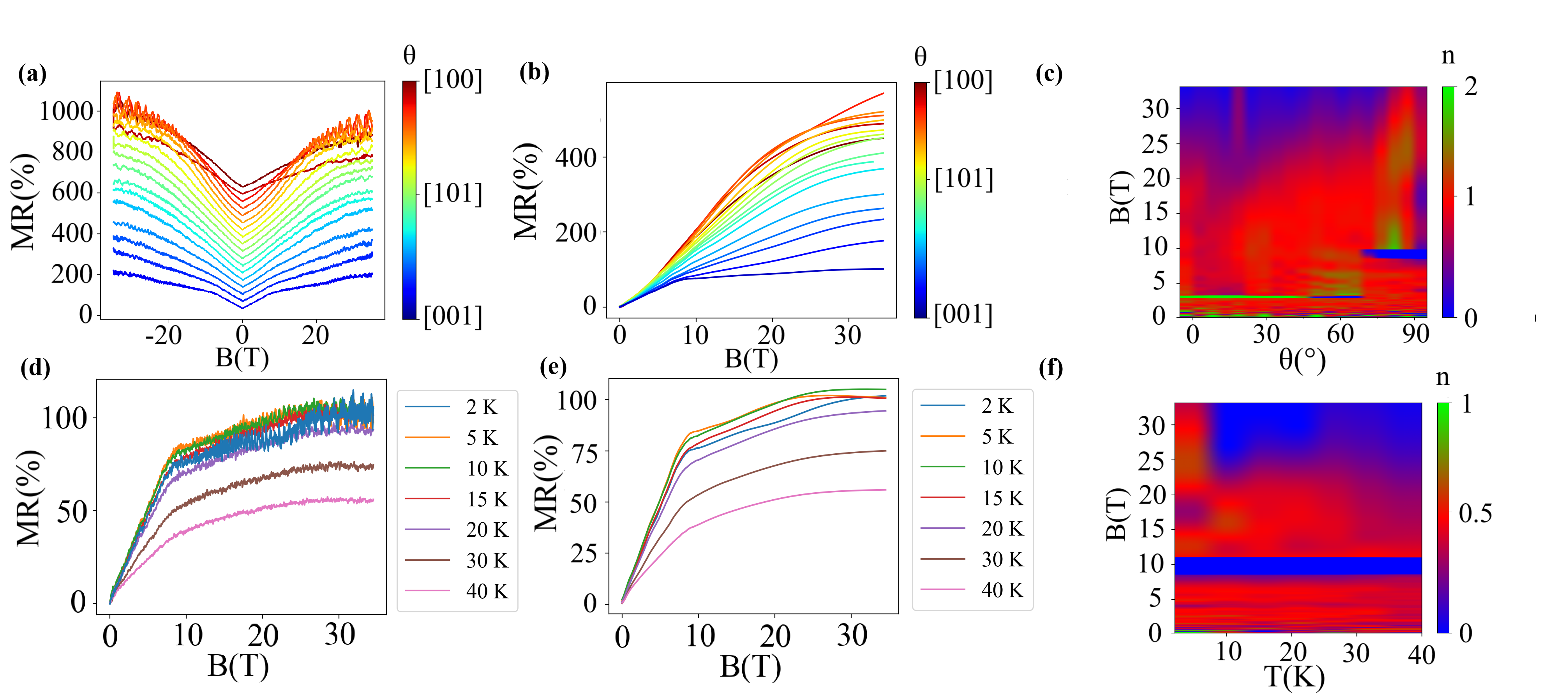}
        \caption{(Color online) (a) Raw magnetoresistance data in configuration C as a function of magnetic field orientation, with an offset for each curve for better visualization. (b) Symmetrized non-oscillatory magnetoresistance data as a function of the magnetic field orientation (after removal of SdH oscillations). (c) Contour plot for the exponent $n(B,T)$ of the power law fit of the non-oscillatory magnetoresistance, obtained as explained in the text, as a function of both magnetic field orientation and the magnetic field intensity. (d) Symmetrized magnetoresistance data as a function of temperature for magnetic field oriented parallel to the \textit{c}-axis. (e) Same as in (d), after removal of SdH oscillations to highlight the non-oscillatory component of MR. (f) Same as in (c), but as a function of temperature.}
        \label{fig:Linear_MR2}
    \end{figure*}

Magnetotransport measurements in configuration C clearly show that the overall non-oscillatory magnetoresistance evolution with field is notably different as the field is tilted from the $a$-axis towards the $c$ axis, compared with rotation within the \textit{a}-\textit{a} plane. For all magnetic field orientations in configuration C  there is an important linear component at low fields for all angles, as evidenced from the V-shaped profile close to zero field in the raw MR data shown in Fig. \ref{fig:Linear_MR2}(a). Furthermore, for magnetic field orientations nearly aligned to the \textit{c}-axis, an inflection field at about $B\approx$ 8 T appears beyond which the magnetoresistance continues to increase linearly but with a lower slope. 

To analyze the field dependence of the non-oscillatory magnetoresistance across different magnetic field orientation, its oscillatory component was removed from the signal prior to performing a power law analysis of the non-oscillatory component. The smoothing procedure to isolate the non-oscillatory background was adjusted according to the magnetic field regime. For magnetic fields above which SdH oscillations are prominent, an interpolation with uniform $1/B$ spacing was performed, and then the magnetoresistance was smoothed using a low-pass filter. For magnetic fields below that value, where the SdH oscillations were not appreciable, the data were interpolated with uniformly spaced $B$ values, and no filter was used.
The threshold field was chosen according to the magnetic field orientation. For field directions nearly aligned with the \textit{a}-axis, the threshold field for the low-pass filter was chosen to be 3 T. In contrast, for field directions close to the \textit{c}-axis, this threshold was 8 T, since there is a clear inflection point close to this magnetic field value. With this, we obtained the non-oscillatory component of magnetoresistance for all orientations in configuration C, as shown in Fig.  \ref{fig:Linear_MR2}(b). 

In order to confirm the linear dependence of magnetoresistance and the presence of two different linear regimes around the \textit{c}-axis, we tracked the evolution of the functional dependence of the magnetoresistance on magnetic field by computing the logarithmic devivative of the MR. This is, assuming a power law dependence of the MR around each field and temperature, of the form $MR=a(B,T)B^{n(B,T)}$, we can find the power law exponent $n(B,T)$ around each field B and temperature T as $n(B,T)=\frac{\text{d}\ln(\text{MR})}{\text{d}\ln(B)}$. For curves between 75° and 90°, this analysis was separated in two regimes, above and below the inflection field of 8 T. For the $B>8$ T regime, a linear dependence was extrapolated to zero field, and the intersection value, $\text{MR}_{8\text{T}\rightarrow 0}$ was subtracted from the magnetoresistance, to calculate the exponent as $n(B,T)=\frac{\text{d}\ln(\text{MR}-\text{MR}_{8\text{T}\rightarrow 0})}{\text{d}\ln(B)}$. The heatmap in Fig. \ref{fig:Linear_MR2}(c) shows the field dependence of $n(B,T)$ for the lowest temperature measured, of $T=1.5$ K, and for all angles across the a-c plane. In this plot, the blue color represents a $n=0$ value for the exponent, the red color represents a linear dependence ($n=1$) and green color represents a quadratic dependence ($n=2$). This plot shows that for all angles and at very low fields there is an important linear ($n=1$, red), component for the non-oscillatory magnetoresistance. For all angles close to the [100] direction, or 0°, this linear component is lost at high fields (above 15 T) and a saturation of the magnetoresistance is evidenced in the $n=0$ exponent value. However, as angle is increased toward 90° (or $c-$direction), the field range for the linear magnetoresistance extends to higher fields, and at 90° the linear magnetoresistance dominates the whole field range that was probed in our measurements.

In addition to the analysis of the non-oscillatory magnetoresistance dependence with magnetic field for different angles, the latter approach was also followed for the dependence in temperature, with the magnetic field parallel to the $c-$ axis ($B$ at 90°). Figs. \ref{fig:Linear_MR2}(d) and (e) show the original symmetrized data, and data after removing the oscillatory component, respectively. For the lowest temperature available, a low frequency component remains, since filtering at such low frequencies is challenging. In spite of this, two linear regimes remain distinguishable up to temperatures as high as $T=20$ K. At even higher temperatures, the high-field linear regime gradually vanishes, and the magnetoresistance rapidly saturates. This behavior is captured in the temperature-dependent heatmap shown in Fig. \ref{fig:Linear_MR2}(c), which illustrates the progressive shift from a linear ($n=1$) to a sublinear ($n<1$) trend in both field regimes as temperature increases.

Linear magnetoresistance has been observed in a variety of quantum systems. To explain this linear dependence, various mechanisms have been considered as routes to this phenomenon. The first of these relies on the quantization of electron orbits in the presence of an external applied magnetic field \cite{Abrikosov_2000}. In this scenario, originally proposed by Abrikosov, a linear magnetoresistance emerges when \textit{all} charge carriers occupy only the lowest Landau level—a condition known as the quantum limit.  However, this can only occur when the FS comprises very small pockets, such that the quantum limit can be reached with magnetic fields that are generated at laboratory (in this case, $B<35$ T \cite{Hu2008,PhysRevB.99.045119}). In \TT{}, although we showed that there are pockets related to frequencies as small as $32$ T, within the range of magnetic fields used for our measurements, its FS is complex and contains various different frequencies of quantum oscillations arising from different pockets, including larger pockets. Thus, the quantum limit explanation is not suitable here.

The second widely cited mechanism attributes linear magnetoresistance to disorder or spatial inhomogeneities in the sample  \cite{Hu2008,10.1063/1.1866642}. In this case, the system is modeled as a random network due to the presence of distorted current paths misaligned with the applied driving voltage, which gives rise to a mix of off-diagonal components in the resistivity tensor, producing an apparent linear dependence on magnetic field.  However, this explanation is also unlikely to apply to \TT{}, as the crystals used in our measurements are high-quality single crystals and exhibit minimal disorder, as evidenced in the large residual resistivity ratio and the presence of well defined quantum oscillations. Therefore, neither the quantum-limit scenario nor inhomogeneity-based models adequately explain the linear magnetoresistance observed in this system.

Other mechanisms proposed to explain  linear magnetoresistance in semimetals include materials with compensated hole and electron carriers. In such systems, linear dependence can dominate over the quadratic term for high magnetic fields \cite{doi:10.1021/acs.nanolett.1c01647,PhysRevB.92.041104,Pan2017}. Although we have identified both hole and electron FS, we do not have concrete evidence that \TT{} is a compensated material. 

Since a secondary linear magnetoresistance domain at high fields as the field orientation becomes parallel to the \textit{c}-axis, the high field linear magnetoresistance domain could be a consequence of the emergence of magnetic breakdown phenomenon close and at this orientation. According to semiclassical models, electronic transport in metals is primarily governed by the scattering rate associated with the mean free path of electrons. However, in a CDW system anisotropic quantum fluctuations can lead to the emergence of \textit{hot spots} in the FS at sections where the FS reconstruction is the strongest \cite{PhysRevLett.29.124}. These hot spots also strongly contribute to the scattering of the charge carriers when they come close to these regions. As a result, the total scattering rate $\tau$ can be modeled as the sum of contributions from `hot spot',  $\tau_{hs}^{-1}$, phonons $\tau_{phonon}^{-1}$, and impurities, $\tau_{i}^{-1}$ scattering rate, or $\tau^{-1}=\tau_{phonon}^{-1}+\tau_{hs}^{-1}+\tau_{i}^{-1}$. When $\tau_{hs}^{-1}$ dominates over other terms, at high fields, it was shown that linear magnetoresistance is obtained, while when phonons and impurities contributions overcome the hot spots contributions, classical quadratic magnetoresistance is recovered, as in quasi-2D CDW rare-earth tritellurides systems \cite{PhysRevB.96.245129}. In the case of \TT{}, the Fermi surface undergoes strong reconstruction due to the CDW transition. If we associate hot spots with regions of enhanced tunneling probability between adjacent FS pockets, then the emergence of magnetic breakdown at these points provides a good explanation for the observed high-field linear magnetoresistance. This interpretation is consistent with both the field-orientation dependence and the field scale at which the secondary linear regime appears.

While magnetic breakdown provides a plausible explanation for the emergence of linear magnetoresistance at high magnetic fields and field orientations near the \textit{c}-axis, a different mechanism must be responsible for the low-field linear behavior. This is evidenced by the presence of a strong linear component in the magnetoresistance at fields close to zero, which is observed consistently across all field orientations in configuration C. For instance, the magnetic breakdown high frequency is not present for orientations around the $a-$direction, but the linear behavior is dominant in the low field regime for these orientations. This strongly suggests that magnetic breakdown is not the relevant mechanism in the low-field regime. Instead, the origin of this low-field linear magnetoresistance may be partially related to the current flow along the \textit{a}-axis in configuration C, which contrasts with that in configuration A, for which the current flow is along the \textit{c}-axis and a quadratic dependence is observed at low fields. Interestingly, at zero magnetic field field the resistivity along $a$, compared to the resistivity along $c$ is not very different, being $\rho_a(B=0)/\rho_c(B=0)\sim 1$ \cite{TADAKI1990227}. This suggests that the linear magnetoresistance, at least for the low-field regime, also has an important contribution of effects driven by magnetic field induced symmetry breaking. In fact, the predicted band structure of \TT{} in both, non-CDW and CDW phases, shows highly degenerate Dirac points close to the Fermi level, for which the presence of a magnetic field can break their degeneracy into different Weyl points. Previous works in other topological semimetals have interpreted the observed linear magnetoresistance as a signature of the chiral anomaly \cite{PhysRevB.93.115414,doi:10.1126/science.aac6089}. In particular, in $Cd_{3}As_{2}$ it has been suggested that the splitting of the Dirac points into two opposite helicity Weyl-points due to the presence of an external applied magnetic field might be related to the observed linear magnetoresistance \cite{PhysRevB.92.081306,Liang2015}. When Dirac points are close to the Fermi level, the shifting (proportional to $g\mu_B B$) of the Dirac points becomes a dominant effect in the transport properties. Thus, Zeeman splitting term surpasses the thermal smearing, releasing some backscattering processes, which do not actually take place at zero field. As previously mentioned, this material is predicted to host highly-degeneracy Dirac points close to the Fermi level. Thus, the topological route for linear magnetoresistance in \TT{} is also feasible and should be considered, although further research is required.

\section{Conclusions}

In summary, our combine high-field magnetotransport and DFT study provides a comprehensive reconstruction of the Fermi surface of TaTe$_4$ in its charge density wave (CDW) phase. We resolve four of the theoretically predicted Fermi surface pockets and find no evidence of non-CDW states, confirming that the bulk electronic structure is fully reconstructed by the CDW. Our measurements reveal a previously unobserved quasi-cylindrical pocket and a high-frequency quantum oscillation arising from magnetic breakdown between reconstructed Fermi surface sheets, from which we estimate a CDW gap of approximately 0.29\,eV, consistent with ARPES-based estimates.

Beyond the Fermi surface mapping, our results uncover a robust linear magnetoresistance that emerges at all field angles when current flows along the $a$-axis. Notably, a second linear regime appears at high magnetic fields near the $c$-axis, which we associate with magnetic breakdown near hot spots in the Fermi surface. However, further research is needed to fully elucidate whether the linear magnetoresistance, particularly in the low-field regime, can be related to the topological character of this compound. Together, these findings establish TaTe$_4$ as a model platform to investigate the interplay between topological features and correlation-driven band reconstruction. They also open new routes for exploring magnetic breakdown physics, anisotropic transport, and topological responses in quasi-one-dimensional semimetals with strong electronic correlations.

%% file: Sections/MaterialsMethods.tex
\section{Materials and methods}

\subsection{Computational details of the electronic structure calculations}

To calculate the electronic band structure and Fermi surface we performed density functional theory (DFT) calculations \cite{PhysRev.136.B864,PhysRev.140.A1133} using the Vienna $ab$-initio simulation package, \textsc{vasp} code (version 5.4.4) \cite{PhysRevB.54.11169,PhysRevB.59.1758}, with the exchange-correlation represented within the generalized gradient approximation
(GGA-PBEsol) parametrization \cite{Perdew2008}. 
The electronic configurations considered in the pseudo-potentials, as valence electrons, are Ta: (5$p^6$6$s^2$5$d^3$, version 07Sep2000) and Te: (2$s^2$5$p^4$, version 08Apr2002).
The energy cut-off for a plane-wave basis is set to 600 eV used to ensure forces convergence of less than 0.001 eV \r{A}\(^{-1}\), and the periodic solution of the crystal was represented by using Bloch states with a Monkhorst-Pack \cite{PhysRevB.13.5188}  \(k\)-point mesh of 13$\times$13$\times$13 in the high-symmetry $P4/mcc$ (SG. 124) phase and scaled
to 9$\times$9$\times$5 in the low-symmetry CDW $P4/ncc$ (SG. 130) phase into the \(2\times2\times3\) supercell representation. The \textsc{Python} library \textsc{PyProcar} \cite{HERATH2020107080} and \textsc{skeaf} \cite{ROURKE2012324} codes were used to analyze the electronic structure of the material and to obtain the predicted extremal surface areas of pockets in reciprocal space.

\subsection{Sample growth}
TaTe\(_{4}\) single crystals were grown using both chemical vapor transport (CVT) and self-flux methods \cite{PhysRevB.102.035125}.

For the self-flux method, Ta and Te powders in a 1:99 ratio were placed inside alumina crucibles and then sealed in clean, evacuated quartz tubes. The powders were heated from room temperature up to 700°C, maintaining this temperature for 12 hours. Subsequently, the temperature decreased to 500 °C within 100 hours. Finally, the crystals of TaTe$_{4}$ were isolated from the melt by centrifuging the mixture immediately after extraction from the furnace.

For the CVT method, Ta and Te powders were mixed in a 1:4 ratio. These were pressed into a pellet and sealed in an evacuated quartz tube. The mixture was heated to 400$^\circ$C and kept for 4 days, then cooled to room temperature. At this point, the tube was opened and the powder was resealed with 2-5 mg/cm$^{3}$ of iodine. The tube was placed in a two-zone tube furnace where the reaction zone was kept at 400$^\circ$C and the transport zone at 370$^\circ$C for 7 days.

TaTe$_{4}$ single crystals were successfully produced using both methods. Both methods resulted in subtle but important morphological differences. The flux method resulted in the typical \textit{c}-axis elongated crystals. The CVT method produced single crystals with a square shape and similar sizes along the \textit{c} and \textit{a} axes. These differences allowed us to obtain distinct current configurations: current flow through the crystalline \textit{c}-axis for the self-flux method obtained crystals, and through the \textit{a}-axis for CVT-grown crystals, which was highly beneficial for the completeness of our study.

\subsection{Sample characterization and transport measurements} 
The chemical composition and crystalline structure of the grown crystals were confirmed using X-ray fluorescence (XRF) and X-ray diffraction (XRD) techniques, with no appreciable differences for flux-grown and CVT grown crystals. Electrical leads were attached to the surface of the crystals after gold evaporation to obtain a good contact resistance, typically below 10 $\Omega$. A standard four-probe connection was employed to measure transport properties using a lock-in amplifier with a low-frequency AC signal. As stated previously, depending on the method used for crystal growth, the current was oriented along the crystalline \textit{c}-axis (self-flux) or \textit{a}-axis (CVT). Magnetoresistance measurements up to 35 T were taken at the DC-field facility of the National High Magnetic Field Laboratory in Tallahassee, Florida.

%% file: Sections/Acknowledgements.tex
\begin{acknowledgments}

A portion of this work was performed at the National High Magnetic Field Laboratory, which is supported by the National Science Foundation Cooperative Agreement No. DMR-2128556* and the State of Florida.
The calculations presented in this article were carried out using the GridUIS-2 experimental testbed, developed in the High Performance and Scientific Computing Center of the Universidad Industrial de Santander (SC3-UIS), development action with the support of the UIS Vicerrectora de Investigación y Extension (VIE-UIS) and several UIS research groups, as well as other funding resources.
A.C.G.C. acknowledge the grant No. 4211 entitled “Búsqueda y estudio de nuevos compuestos antiperovskitas laminares con respuesta termoeléctrica mejorada para su uso en nuevas energías limpias” supported by Vicerrectoría de Investigaciones y Extensión, VIE--UIS. E. R. R and P.G.G acknowledges financial support from the project ‘Ampliación del uso de la mecánica cuántica desde el punto de vista experimental y su relación con la teoría, generando desarrollos en tecnologías cuánticas útiles para metrología y computación cuántica a nivel nacional’, BPIN 2022000100133, funded by the General Royalties System (SGR) of MINCIENCIAS, Government of Colombia. 
E.H.V., P.G.G. and J.A.G. acknowledge the QUASURF project (SI4/PJI/2024-00062) funded by the Comunidad de Madrid through the agreement to promote and encourage research and technology transfer at the Universidad Autónoma de Madrid. 
P.G.G. acknowledges support of Facultad de Ciencias, Universidad de Los Andes, through project INV-2023-176-2931.
\end{acknowledgments}

\section*{Author Contributions}
D.S-V., J.R-C and P.G-G. did the single crystal sinthesis. D.S-V., J.R-C, P.G-G., E.H-V., J.A.G. and S.C took the high field transport data. B.U., R.G-H and A.C.G-C. did the DFT calculations. All authors participated in the data analysis, discussion and manuscript writing and revision.

\section*{Data availability}
Data used and produced in the development of this work will be made available from the author upon reasonable request.

%% file: Sections/SupplementaryInfo.tex
\section*{Supplementary Information}
\setcounter{section}{0}
\setcounter{figure}{0} 
\makeatletter 
\renewcommand{\thefigure}{S\@arabic\c@figure}
\renewcommand{\thesection}{S.\Roman{section}}
\renewcommand{\thesubsection}{S.\Roman{section}.\Alph{subsection}}
\makeatother
\section{Effective masses}\label{appendix_a}

To estimate the contribution of
charge carriers to the electronic properties of TaTe\(_{4}\), the Liftshitz-Kosevich (LK) formula for the oscillatory magnetoresistance is used\cite{shoenberg_1984}:

\begin{align}
\begin{split}
\frac{\tilde\rho(B)-\rho_{0}}{\rho_{0}}=\sum_{i} C_{i}\{\text{exp}\left(\frac{-14.7\text{ T/K}(m_{i}^{cyc}/m_{e})T_{D,i}}{B}\right)\}\\
\times\{\frac{T/B}{\text{sinh}[14.7\text{ T/K}(m_{i}^{cyc}/m_{e})T/B]}\}\\
\times\text{cos}[2\pi(\frac{F_{i}}{B}+\phi_{i})]
\end{split}
\end{align} 

where the sum is over all contributing frequencies, \(m_{e}\) represents the electron mass, \(T_{D}\) the Dingle temperature, which relates to the free mean path of charge carriers and \(m_{i}^{cycl}\) the cyclotron effective mass of the frequency \(F_{i}\). The effective masses for the pockets can be obtained by fitting the  evolution of the oscillatory magnetoresistance with temperature using the previous formula for a fixed angle. In our case, the effective masses were estimated by filtering the oscillatory signal of each pocket with a band-pass filter to select a single frequency, resulting in a better fit to our data than by performing the fit on the sum of all frequencies with the whole LK formula. Dingle temperatures were also estimated, but the error bars for this parameter were in the range of the temperatures obtained.

\begin{figure*}[!t]
        \centering
        \includegraphics[width=\textwidth]{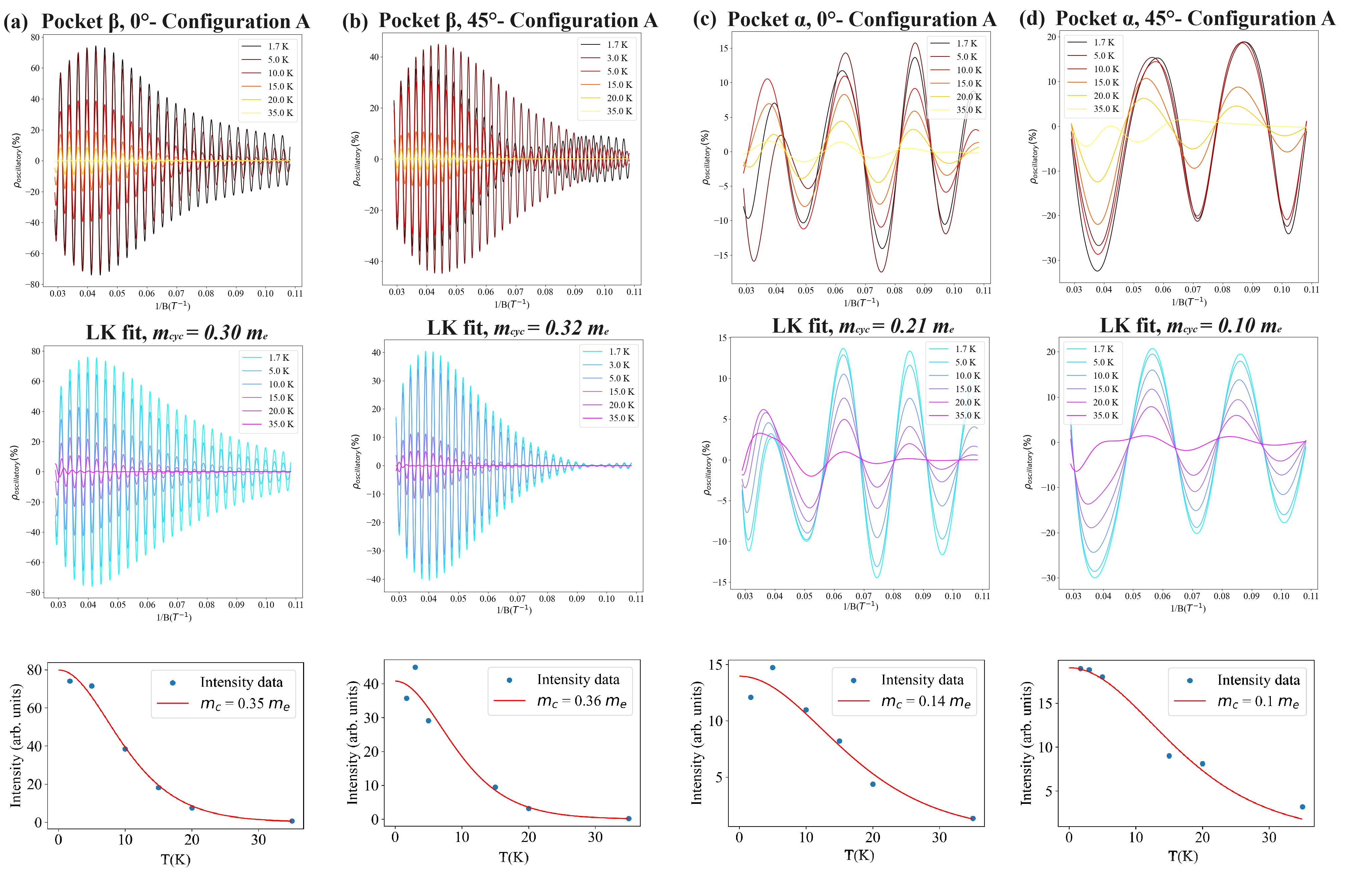}
        \caption{(Color online) Temperature dependence of the amplitude of the oscillating component of magnetoresistance for TaTe\(_{4}\) for some dominant pockets and magnetic field orientations in Configuration A. Columns (a) and (b) show the oscillatory magnetoresistance data and corresponding LK fits for pocket $\beta$  at 0° and 45° respectively, while columns (c) and  (d) are structured similarly but for pocket $\alpha$. The top row displays the oscillatory data filtered for each pocket at different magnetic fields and temperatures, the middle row shows the LK fit and the cyclotron mass associated to that fit of the data at the top row, and the bottom row shows the LK fit for a single magnetic field intensity and different temperatures. The middle column plots show the fits of
the data to the LK formula of the selected pocket.
The values of cyclotron effective masses are obtained from this fit. The right column shows the LK for a fixed magnetic field value, at different temperatures.}
        \label{fig:Effective_masses}
    \end{figure*}

Using this approach, and following the magnetic field orientations defined for configuration A in previous sections, we obtained \(m^{cyc}_{\beta-0°}=(0.30\pm 0.03)m_{e}\) for the lower frequency branch of pocket \(\beta\) at 0\(^{\circ}\)  and \(m^{cyc}_{\beta-45°}=(0.32\pm 0.03) m_{e}\) at 45\(^{\circ}\) in configuration A (figs. \ref{fig:Effective_masses}(a,b)), which suggests that the mass of charge carriers in these pockets is highly isotropic along this rotation plane.

Since pocket \(\beta\) resembles an ellipsoid according to DFT predictions, it is interesting to address whether the pocket can be described following a perfect parabolic dispersion model for energy and an ideal ellipsoidal model \cite{PhysRevB.94.195141}. According to the perfect ellipsoidal model, $\frac{m^{cyc}(\theta)}{m^{cyc}_{\perp}}=\sqrt{\frac{K}{(K-1)\cos^{2}\theta+1}}$, where $K=\left(\frac{f_{max}}{f_{min}}\right)^{2}$ is the pocket anisotropy. For pocket \(\beta\), \(m^{cyc}_{\beta-45°})/m^{cyc}_{\beta-0°}=1.07 \pm 0.21\), and \(K=(f_{max}/f_{min})^{2}=(482/333)^{2}=2.09\). Since \(m^{cyc}_{\beta-45°}/m^{cyc}_{\beta-0°}\approx \sqrt{2K/(K+1)}\) by taking into account the error bars, this pocket approximately follows the perfect ellipsoidal model.

On the other hand, pocket \(\alpha\) seems to be anisotropic in terms of effective masses and shape (figs. \ref{fig:Effective_masses}(c,d)). The effective mass of the charge carriers smoothly changes with angle for pocket \(\alpha\), with \(m^{cyc}_{\alpha-45°}=(0.21 \pm 0.03)m_{e}\) at 0\(^{\circ}\) and \(m^{cycl}_{\alpha-0°}=(0.10 \pm 0.02) m_{e}\) at 45\(^{\circ}\) in configuration A, indicating a high mass anisotropy for this pocket. Additionally, figure \ref{fig:Effective_masses}(c,ii) shows that the oscillation frequency of pocket \(\alpha\) at 0\(^{\circ}\) is \(f_{min}=\)44 T, whereas at 45\(^{\circ}\) this value is \(f_{45}=\)32 T. Then, \(K=(f_{45}/f_{min})^{2}=1.89\), which is similar to the effective mass anisotropy, \(m^{cyc}_{\alpha-45°}/m^{cyc}_{\alpha-0°}=2.10 \pm 0.72\).

In contrast to pockets \(\alpha\) and \(\beta\), the frequencies associated with pocket \(\zeta\) vanished rapidly as temperature increased. Since the LK formula depends on both magnetic field and temperature, it could not be used to estimate effective masses for this pocket. 